\documentclass[A4]{aa}
\usepackage{aalongtable}
\usepackage{graphicx}
\usepackage{natbib}
\usepackage[dvips]{color}
\begin{document}
\title{3D-radiation hydro simulations of disk-planet interactions}
\subtitle{I. Numerical algorithm and test cases}
\titlerunning{3D-radiation hydro simulations of disk-planet interactions}

\authorrunning{Klahr and Kley}
\author{H. Klahr\inst{1,2}
\and
W. Kley\inst{1}}
\offprints{H. Klahr}
\institute{Universit\"at T\"ubingen, 
Institut f\"ur Astronomie und Astrophysik, 
Abt. Computational Physics,
Auf der Mor\-gen\-stelle 10, D-72076 T\"ubingen, Germany
\email{wilhelm.kley@uni-tuebingen.de}
\\
\and
Max-Planck-Institut f\"ur Astronomie,
K\"onigstuhl 17, D-69117 Heidelberg\\
\email{klahr@mpia.de}
}
\date{Received  / Accepted }
\abstract{We study the evolution of an embedded protoplanet in a circumstellar 
disk using the 3D-Radiation Hydro code TRAMP, and treat
the thermodynamics of the gas properly in three dimensions.
The primary interest of this work lies in the 
demonstration and testing of the numerical method. 
We show how far numerical parameters can influence
the simulations of gap opening. We study
a standard reference model under various numerical approximations.
Then we compare the commonly used locally isothermal approximation
to the radiation hydro simulation using an
equation for the internal energy.
Models with different treatments of the mass accretion process
are compared. Often mass accumulates in the Roche lobe of the planet
creating a hydrostatic atmosphere around the planet.
The gravitational torques induced by the spiral pattern of the disk
onto the planet are not strongly affected in the average 
magnitude, but the short time scale fluctuations are
stronger in the radiation hydro models.

   An interesting result of this work lies in the analysis of the temperature
structure around the planet. The most striking effect of treating the thermodynamics properly
is the formation of a hot pressure--supported bubble around the planet 
with a pressure scale height of $H/R \approx 0.5$ rather than a thin Keplerian 
circumplanetary accretion disk.
We also observe an outflow of gas above and below the planet during the gap opening phase.
}
\maketitle
\section{Introduction}
During their formation process, massive, gaseous planets are
believed to undergo a phase of evolution where they are still
embedded in the protoplanetary disk. The gravitational influence of
the planet onto the ambient disk leads to such features as spiral arms
and, for planets sufficiently massive, an annular gap at the planetary
radius. The back-reaction of the perturbed disk onto the planet
generates torques, which induce a change of the orbital elements
(semi-major axis and eccentricity) of the planet. 
The linear phase for small planetary masses has been studied 
for two--dimensional flat disks by \citet{1980ApJ...241..425G} and
\citet{1997Icar..126..261W}, and more recently in three dimensions
by \citet{2002ApJ...565.1257T}. Analytical formulae
for the (inward) radial migration rate have been obtained.
These linear (local)
analyses have been supplemented
by a variety of numerical simulations that also
consider the global disk and larger planetary masses.
Two--dimensional flat disks with planets were modeled by
\citet{1999ApJ...514..344B, 1999MNRAS.303..696K} and \citet{1999ApJ...526.1001L}
applying a locally isothermal assumption for the radial
temperature distribution.
Later, global three--dimensional simulations of embedded planets were performed
on equidistant grids \citep{2001ApJ...547..457K}, with nested
grids \citep{2003ApJ...586..540D} and by applying local grid refinement
\citep{2003MNRAS.341..213B}. They confirmed the linear results for very small
planets as well as the estimates for gap opening large planets.
In these simulations the planet is held on a fixed circular orbit
and the mass accretion and migration rate are estimated from the obtained
density distribution.
Simulations with a moving single planet 
\citep{2000MNRAS.318...18N} are again in agreement with the inward
migration scenario.

These hydrodynamical simulations have been extended recently to
full 3D-MHD calculations though neglecting vertical gravity
\citep{2003MNRAS.339..983P, 2003MNRAS.339..993N,
2003ApJ...589..543W}. Here, the turbulent state of the disk
is modeled directly, and the migration rate is found to fluctuate
strongly around a mean value.

All the above mentioned simulations treat the disk as isothermal locally, 
although polytropic disks
have been considered in \citet{1999MNRAS.303..696K}. Recently,
\citet{2003ApJ...599..548D} added thermodynamical effects in
a two--dimensional model. They assume that the dissipation
produced in the disk (by viscosity) is radiated locally, while
the equation of state includes dissociation and ionization
effects. 
Using this method in addition to nested grid refinement around the planet
it also has been possible for the first time to obtain an estimate of the
two--dimensional temperature structure in the disk's midplane in the vicinity
of the planet. 
However, these simulations still suffer from their two--dimensionality. 

In the present paper we  extend the radiation hydrodynamical planet--disk calculations to
three dimensions, by applying a flux--limited diffusion approximation 
for the radiation.
This allows us to study directly the dynamical influence of the planetary
accretion luminosity and to determine the three--dimensional temperature structure in the vicinity the planet.

Here, we focus on numerical issues, and study the effects
of the radiation transport, of resolution and mass accretion.
By introducing radiation and opacities the model is 
no longer scale--free, as it is in the isothermal case. 
Hence, there
is a difference,
whether the planet is located at 5 AU or 0.5 AU, say.
In addition to influencing the gravitational torques acting on the planet,
the disk mass now has an effect
on the temperature structure as well, because it provides the optical depth.

We find that for the test parameters (1 Jupiter mass at 5 AU) and the low numerical 
resolution that we apply, there is no significant acceleration or breaking of the
mean radial drift in comparison to the isothermal approximation.
However, the fluctuations in the radial drift appear an order of magnitude larger
in the radiative models with realistic accretion than in the isothermal models.

We find the circumplanetary ``accretion disk'' to be rather thick, i.e. the pressure
scale height is $H/R \approx 0.5$ in our models, which is even thicker than the results from 2D simulations \citep{2003ApJ...599..548D}. 
The deviation from a Keplerian
profile in this ``disk'' is about $50\%$, which also indicates the strong support from the thermal pressure.
Under such conditions the thin disk approximation breaks down and one has to deal with 
a circumplanetary cloud filling a sizable fraction of the Roche lobe rather than a disk.

The detailed modeling of the observational appearance of 
a planet interacting with a disk and a detailed parameter
study are left to follow up studies.

In the next section we describe our model setup, the boundary and
initial conditions. In section 3 we present an analysis of our models,
and we conclude in section 4. 
\section{Model characterization}
We consider a Jupiter sized protoplanet embedded in a protoplanetary
disk in three dimensions, where we use a spherical polar
coordinate system ($r, \theta, \varphi$). The central star
lies at the origin of the coordinate system and the disk
midplane coincides with the { $\theta=0$} plane.
We work in a coordinate frame corotating with the planet, and use
a special formulation avoiding explicit source terms
to conserve angular momentum locally and globally, see
\citet{1998A&A...338L..37K}.
\subsection{Physical setup}
All models we consider in this paper have a radial extent from 
$1.25$AU to $20$AU, where a Jupiter mass planet is located at 5 AU.
The vertical extent is from $-10^{\circ}$ below the
midplane to $+10^{\circ}$ above the midplane of the disk. 
In the azimuthal $\varphi$-direction we consider a complete annulus
from 0 to $2\pi$.
The initial temperature distribution is proportional to $r^{-1}$
and ``vertically'' (i.e.\ in the polar direction) constant.
This temperature is chosen to reproduce a local pressure scale 
height $H/r = 0.05$ for all radii.
The actual vertical extent of the computational domain is then
approximately $\pm 3.5 \times H/r$.
The initial density distribution is radially proportional to
$r^{-1.5}$ with a midplane density at 5 AU of
$\rho({\rm 5 AU},\theta = 0) = 10^{-11}$ g cm$^{-3}$. 
This corresponds approximately to
a local surface density $\Sigma = 2 H \rho = 75~[{\rm g\, cm}^{-2}]$ at 5 AU, 
i.e.\ the location of the planet. 
This is similar to the value used by 
\citet{1999ApJ...514..344B, 1999MNRAS.303..696K, 1999ApJ...526.1001L}.

Due to the assumed local vertical isothermal disk as initial setup,
the vertical density structure is given to first order 
by a Gaussian:
\begin{equation}
  \rho(r,\theta)= \rho(r,0) \, e^{ - \theta^2\left(\frac{r}{H} \right)^{2}}.
\end{equation} 
The vertical $v_\theta$ and radial $v_r$ velocities vanish initially. 
The initial azimuthal velocity is
given by the equilibrium of gravity, centrifugal acceleration and the
radial pressure gradient. 
This is not an exact solution
because the Gaussian density distribution is only valid in the 
thin disk approximation. 
Thus, as we resolve the disk out to
$\pm 10^{\circ}$ a small discrepancy appears.
Nevertheless, the model is only slightly out of equilibrium and
will relax quickly into a more stable state, which deviates only
marginally from the initial state.

In the models we do not use any explicit physical disk viscosity
and model the ideal hydrodynamical equations. 
However, we use an artificial bulk viscosity in the form of
\citet{vonNeumann1950}, see \citet{1992ApJS...80..753S}.
Applying an explicit viscosity, e.g.\ $\alpha$-viscosity would have 
introduced one further free parameter, which we seek to avoid.
Although MHD turbulence is a good candidate to produce an $\alpha$-like viscosity \citep{1973A&A....24..337S},
it is not obvious that $\alpha$ would be a constant in time and space 
\citep[see e.g.\ ][]{2005MNRAS.362..361W}. It is also unclear in how far the magneto--rotational
instability would operate in the gap region and in the circumplanetary 
bubble.

A second reason for not using an explicit viscosity is that it would have led
to additional heating on top of the adiabatic contraction of the circumplanetary 
bubble, thus making it more difficult to identify the importance of the heating by the accretion onto 
the planet.

%
\subsection{Gravitational potential}
{ The Roche lobe around the planet determines the region where mass 
is gravitationally bound to the planet. The
Roche lobe is pear--shaped with a sharp edge at the side 
towards the star and thus is usually 
approximated by the Hill sphere. This sphere is centered at the location
of the planet and completely includes the Roche lobe.
We will usually refer to the Roche lobe and its approximate
size given by the Hill radius.}

To calculate the potential due to a planet of mass $M_p$ at the
radius $r_p$ in the midplane of the accretion disk, 
we determine 
the radius $a_\mathrm{h}$ of the Hill sphere around the planet, which is a function of
the reduced mass $\mu = \frac{M_p}{M_\odot + M_p}$ of the system
\begin{equation}
a_\mathrm{h} = r_\mathrm{p} \left(\frac{\mu}{3}\right)^{\frac{1}{3}}.
\end{equation}
For distances from the planet larger than a critical fraction $q_\mathrm{g}$ 
of the Hill radius $a_\mathrm{h}$ we use the exact gravitational potential
$\Phi$ as the sum of stellar $\Phi_\odot$ and planetary $\Phi_p$ potential
\begin{equation}
\Phi \, = \, \Phi_\odot + \Phi_{p} \,  = - \, \frac{M_\odot G}{r} 
      -  \frac{M_\mathrm{p} G}{\sqrt{({\bf r}-{\bf r}_{p})^2}},
\end{equation}
where ${\bf r}$ is the radius vector pointing from the star to a specific location
in the disk.
For closer distances we use a cubic expansion which is 
2$^{\rm nd }$order accurate at a critical distance $a_\mathrm{g} = q_\mathrm{g} \, a_\mathrm{h}$ 
(i.e. it agrees with the outer potential in its 1$^{\rm st}$ and 2$^{\rm nd}$ derivative)
and then becomes shallower towards
the center with a vanishing gradient at the planet location:
\begin{equation}
\Phi_p^* = 
   - M_p G \left[\frac{\left(r-r_p\right)^3}{a_\mathrm{g}^4} 
     - 2\frac{\left(r-r_p\right)^2}{a_\mathrm{g}^3} + \frac{2}{a_\mathrm{g}}  \right],
\end{equation}
The benefit of such a treatment for the smoothing is that the
potential is accurate down to the critical distance, whereas using the
more classical approach
\begin{equation}
\Phi_p^+ =  \frac{M_p G}{\sqrt{\left({\bf r}-{\bf r}_p\right)^2+a^2_{g}}}  
\end{equation} 
leads to a significant underestimation of the potential depth already 
at $a_\mathrm{g}$.

In our models we test the influence the parameter $q_\mathrm{g}$ has on the
mass of gas that accumulates in the { Roche lobe} of the planet.
\subsection{Accretion}
Accretion is treated as in \citet{1999MNRAS.303..696K}, where
a certain fraction of mass is taken out per time step within a 
given accretion radius $a_\mathrm{acc} = q_\mathrm{acc} a_\mathrm{h}$ around the planet. 
Here, we take this fraction $q_\mathrm{acc}$ of the { Hill radius} as 
a free parameter and study its influence on the gas density and temperature in the
vicinity of the planet.
We believe that our models that treat accretion and conserve the total
energy (DR and DR4) have the highest
relevance for the temperature structure of the circumplanetary cloud.
These models yield a lower estimate for the temperatures, because
using a deeper potential might still increase the temperatures.
We also measure the mass contained in the Roche lobe and the
mass change in the Roche lobe per time unit, which is a
good tool to show that the planet is accreting at a steady rate and that the system is in equilibrium, i.e.\ no mass is piling up in the Roche lobe any longer. 
%
\subsection{Numerical scheme}
We use the TRAMP hydro code as described in \citet{1999ApJ...514..325K}.
TRAMP is a versatile 3D radiation hydro package that can use various advection schemes. 
For the purpose of this paper we apply a monotonic transport scheme 
\citep{1977JCoPh..23..276V}.
The radiation transport in TRAMP is treated in the
flux--limited diffusion approximation using the flux 
limiters of \citet{1981ApJ...248..321L}, see also \citet{1989A&A...208...98K},
and dust opacities typical for the solar nebula \citep{1994ApJ...427..987B}.
The scheme to treat internal energy and radiation energy density
is as in \citet{1996ApJ...461..933K}.
Adding the equations for internal and radiation energy and assuming that
the latter is much smaller than the internal energy, which is certainly
a very good approximation for protostellar disks we arrive at the
following evolution equation for the internal energy
\begin{equation}
\label{eq:energy}
\partial_t e + \nabla \cdot \left({\bf u} e \right) = - p \nabla \cdot {\bf u} + \Phi
  - \nabla \cdot {\bf F} 
\end{equation}
where ${\bf u}$ is the velocity vector, $p$ the thermal pressure, $\Phi$
is the dissipation by viscosity. Here we use only artificial viscosity in
shocks to achieve the proper entropy generation but no explicit physical viscosity
in the smooth flow. Nevertheless, tensor viscosity is
fully implemented in TRAMP and we can use it in future simulations of gap opening.
For this paper containing test simulations we avoid the additional
effects of viscous heating in order to study more carefully the accretional heating
in the vicinity of the planet.
 
The radiation flux ${\bf F}$ is calculated in flux--limited diffusion approximation
\begin{equation}
 {\bf F} = \, - \, \frac{\lambda c}{\rho \kappa} \nabla E_{\rm R}
\end{equation}
with the Rosseland mean opacity $\kappa$ and the speed of light $c$.
The flux limiter $\lambda$ is defined according to \citet{1981ApJ...248..321L}.
Here we assume local thermal equilibrium and set
\begin{equation}
\label{eq:lte}
    E_{\rm R} = a T^4 .
\end{equation}
Numerically, the radiative diffusion part is solved separately
as part of an operator splitting technique.
The diffusion equation for $E_{\rm R}$ 
\begin{equation}
  \partial_t E_{\rm R}  = - \nabla {\bf F} = \nabla \, \frac{\lambda c}{\rho \kappa} \nabla E_{\rm R}
\end{equation}
is solved implicitly to avoid time step limitations using e.g.\
the standard SOR ({\it successive over--relaxation}) scheme.
The new temperature is then calculated from Eq.~(\ref{eq:lte}).

The code has been extensively tested in the context of accretion disks \citep{1999ApJ...514..325K}.
The new part that is implemented to the code is the inclusion and 
modification of the gravitational potential of the planet as described above.
\subsection{Boundary conditions}
The boundary conditions are chosen to minimize the effects of wave
reflections at the boundaries. In the polar and radial direction
these are outflow conditions not allowing for inflow.
Thus, for pressure, density and orthogonal velocities, we apply a zero gradient condition, while
the normal velocity components have a vanishing gradient only if
the direction of the flow is outward of the computational domain.
For inward directed flow the normal velocities components are
set to zero. For the angular velocity we assume a Keplerian flow.
No sponge layer as in \citet{2003ApJ...582..869K} is used.
In the azimuthal direction we apply periodic boundary conditions.
\section{The models}\label{sec:models}
The purpose of all the different models using the same physical parameters,
i.e.\ stellar, planetary and disk mass as well as planet location, is to study
the influence of the numerical treatment of the problem and to find the 
numerical parameters suited to study the gap opening as { realistically} as possible.

In Table~(\ref{tab1}) we give an overview of all the models we compare in the following.
The second column (grid) states the numerical resolution ($n_r, n_\vartheta, n_\varphi$) of the grid.
There are low resolution (100$\times$20$\times$200) models
with a logarithmic radial spacing and equidistant in the azimuthal and vertical direction.
We also calculated some high resolution models, where the grid is in all
dimensions logarithmically concentrated (indicated by $lc$) towards the
location of the planet, which results in a resolution in the Roche lobe up to four times higher
than in the low resolution case. In these models there are about 2400 grid points within
the { Roche lobe}. We tested the
influence of the logarithmically distorted high resolution grid on a
disk without a perturbing planet. 
Despite the grid refinement around the planet and quite distorted 
grid cells towards the edges of the grid, there are
no unstable growing modes.

Some models are calculated in the standard
locally isothermal approximation (EOS = iso), some treat the internal energy equation
plus flux--limited radiation transport (EOS = rad).
We also state the parameters for the smoothing of the gravitational potential
$q_\mathrm{g}$ and for the mass accretion onto the planet $q_\mathrm{acc}$.
Some models start with no gap initially, 
others use an axisymmetric initial model with a gap (ini = G). 

   After briefly mentioning a test model for the non-equidistant high resolution grid,
we will go through our models in the following order:
first isothermal models, then radiative models, followed by non-accreting models,
and finally accreting and energy conserving models.
\begin{figure}[ht]
\resizebox{\hsize}{!}
{\includegraphics{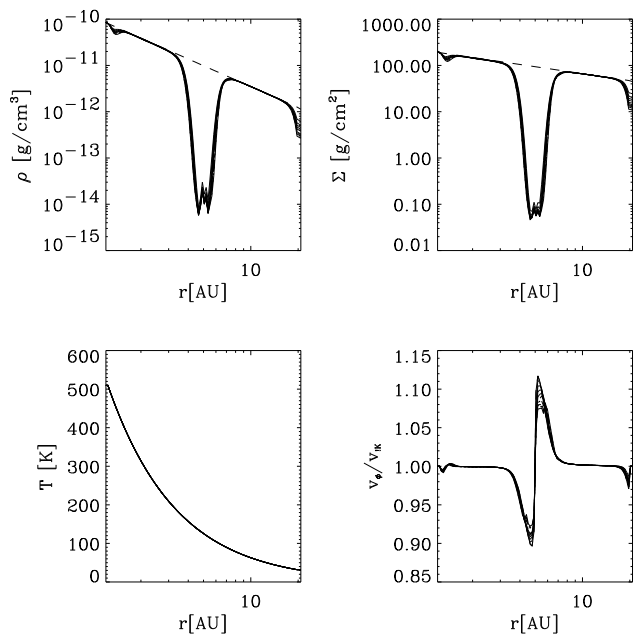}}
\caption{\label{fig:modCI_pstats.ps} 
Model CI: Snapshots taken every 26 orbits from t=60 to t=320 orbits, which is after the convergence of model CI. 
The radial slices of the density, temperature and azimuthal velocity are taken in the midplane at the location of the planet. 
Model CI defines the initial values for all  the models that start off with an axisymmetric gap.
}
\end{figure}
\subsection{Isothermal Models}
\subsubsection{The axisymmetric model CI}
\begin{figure}[ht]
\resizebox{0.98\linewidth}{!}{%
\includegraphics{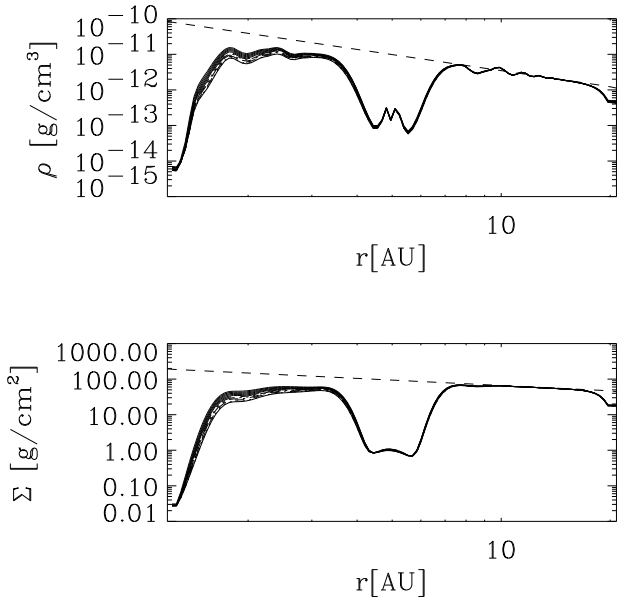}}
\caption{\label{fig:modCDI_pstats.ps} 
Model CDI: Snapshots taken every 2 orbits from t=164 to t=184 orbits. 
The radial slices of the density are taken in the midplane at the location of the planet. The surface density is azimuthally averaged.
}
\end{figure}
\begin{figure}
\resizebox{\hsize}{!}
{\includegraphics{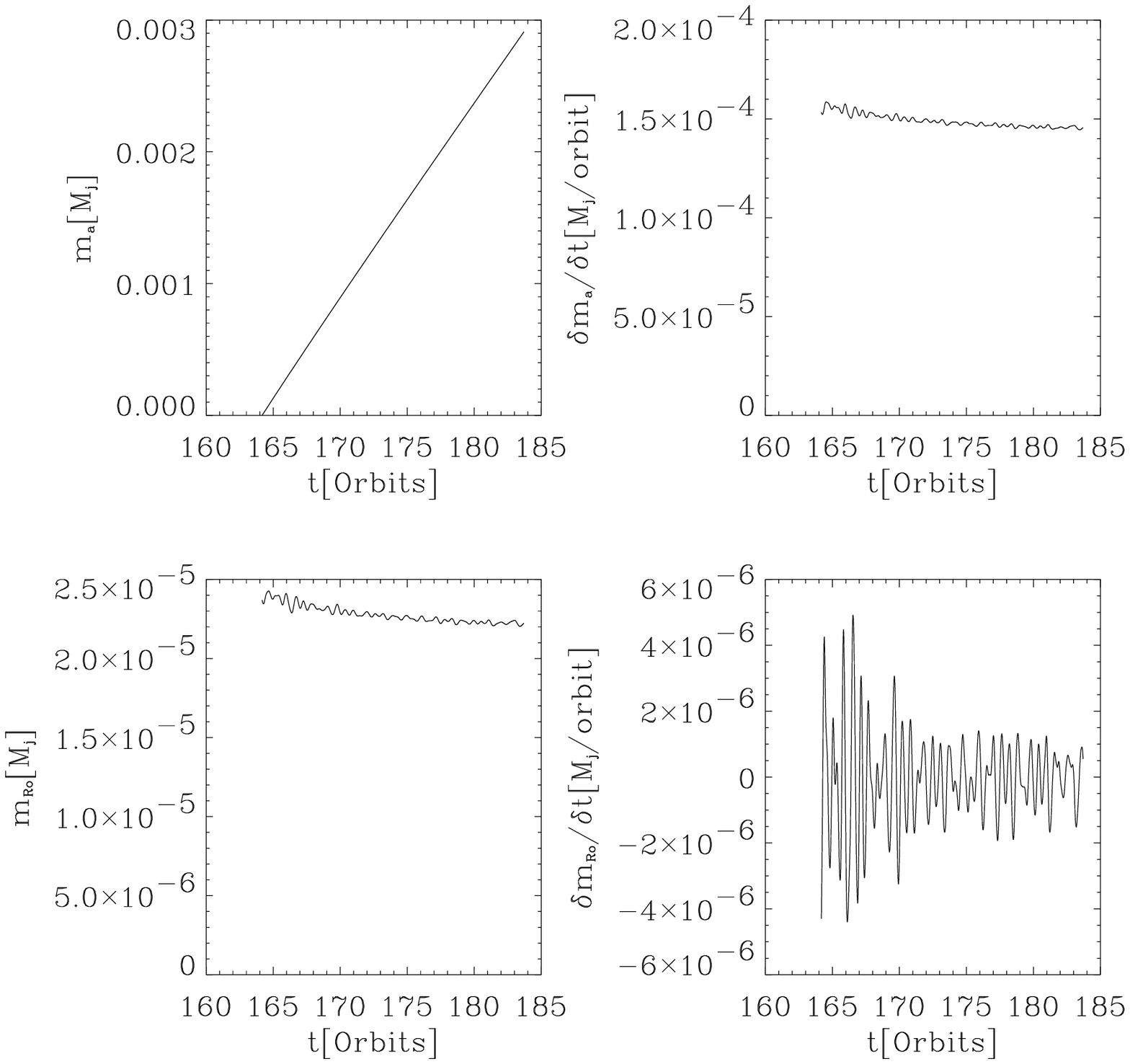}}
\caption{\label{fig:modCDI_planetmass2.ps} 
Model CDI: (upper row) accreted mass $m_a$ and accretion rate 
onto the planet $\dot m_a$ over time; (lower row) evolution of the mass contained in the Roche lobe $m_{\rm Ro}$ and
mass change in the Roche lobe $\dot m_{\rm Ro}$ in units of Jupiter masses and orbital periods of the planet.
}
\end{figure}
\begin{figure}
\resizebox{\hsize}{!}
{\includegraphics{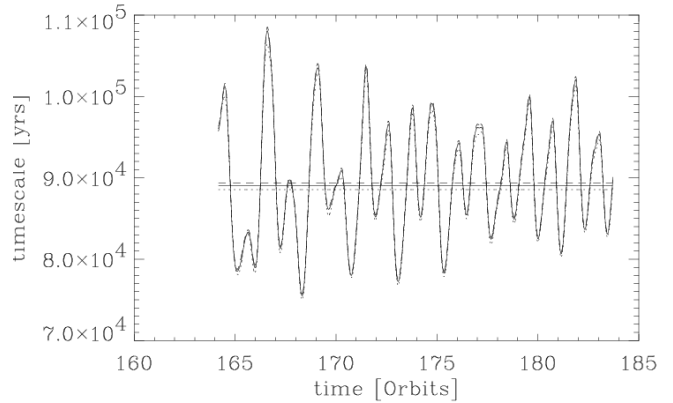}}
\caption{\label{fig:modCDI_torques.ps} 
Model CDI: Evolution of planet migration rate in terms
of migration time. The solid curve is based on the 
torques excluding the Roche lobe and the solid horizontal
line is the time average. In addition we plot the torques based
on an exclusion of 1.5 and 0.5 Hill radii (dotted and dashed lines respectively).
In this particular model the gap is very clean, such that all three torques
mostly coincide.
}
\end{figure}
This model uses only one grid-cell in the azimuthal direction but it
includes the full potential of the
planet, and mass is taken out of the grid from
the center of the Roche lobe.
It is used to
run a model quickly into an initial state with a gap, which
forms here not due to the gravitational torques by the planet but rather
by the perturbation in gravity as it would result from a massive ring. However, as we shall see
the obtained gap size and shape are in reasonable agreement with the
realistic 3D case.

   In Fig.~\ref{fig:modCI_pstats.ps} we introduce the
kind of plot that we will subsequently use for some of the following models.
It shows quantities measured in the midplane at the azimuthal position
of the planet. Thus, it is one--dimensional data. We plot
10 snapshots, which are taken at equal time intervals,
e.g.\ in the case of Fig.~\ref{fig:modCI_pstats.ps} every 26 orbits from t=60 to t=320 orbits.
First we plot the density $\rho$. One clearly sees how the gap forms
and how stable it is. Most of the disk is still at the
initial values, which are given by the dashed line.
The next plot shows the azimuthally
averaged surface density $\Sigma$. 
This value is slightly higher than $75[{\rm g}{\rm cm}^{-2}]$ at 5 AU,
because here we integrate over
the total height of the computational domain and not only over $\pm H$. 
The temperature $T$ in the third plot does not fluctuate as this run is
isothermal. 

The deviation of the local azimuthal 
velocity ($v_\phi/v_K$) from the Keplerian value is significant only in the vicinity
of the ``planet''. The shape of the rotational profile is determined by the
gradient in the surface density and thus has no physical relevance for
the real gap case where the gravitational torques are determining the shape of the 
gap. We also have to mention that this artificial gap structure is not unique,
but there is a family of solutions, where either the surface density profile
or the azimuthal velocity profile is a free parameter.
Still this method generates a gap structure that is not too unrealistic
to serve as an initial model for our simulations.

{ The stationary state of model CI}
is given by the balance of centrifugal forces, radial pressure gradients and
the attraction of the combined stellar and ring potential,
is used as the initial setup in all of the subsequent models that
include a C in their names (see Tab.~\ref{tab1}).
%
Models without a C are started from
the same initial condition as model CI is started, which is a disk without a gap.
\subsubsection{Model CDI}
This is the standard isothermal three--dimensional model. 
Mass is taken out of the Roche lobe. 
Thus, the model is directly comparable to the ones in \citet{2001ApJ...547..457K}.
The radial midplane profiles (see Fig.~\ref{fig:modCDI_pstats.ps})
are measured at equal times between $164$ and $184$ orbits.
The profile only changes very slowly.
In the inner part of the disk one recognizes a decrease
in density, which is related to two effects: First, 
the accretion process triggered
by the tidal forces of the planet, which 'drives' the material through
the inner boundary simulating accretion onto the star. 
This is possible because we use non-reflecting boundary conditions
at the radial boundaries, which allow for outflow but not for inflow.
As an effect also the most outer grid cells loose a lot of mass, 
which shows at most in the high resolution models, e.g.\ model CD4I.

The second reason for the mass depletion is the accretion process 
onto the planet, which occurs from both sides, from the inner as well
as from the outer part of the disk. But the mass reservoir in the outer part
of the disk is much larger, thus the decrease in mass is not
visible over the span of the simulation.
In comparison to the simulations by \citet{1999ApJ...514..344B}, \citet{1999MNRAS.303..696K}, and
\citet{1999ApJ...526.1001L} the gap in the surface density is shallower.
Those simulations used explicit tensor viscosity and started already
with a wider gap than we used. So in our case it will take many hundreds
of orbits before the gap will reach a comparable depth. In terms
of computational time such a run would take several months of 
computation time on our fastest available workstations.
The code in its present version is not parallelized, due
to the implicit solver of the radiation part.
In the future we will consider runs with a much longer evolution time.

Plotting the mass accreted onto the planet and the mass accumulated in
the Roche lobe (see Fig.~\ref{fig:modCDI_planetmass2.ps}) also shows
that we are only slowly approaching a steady state of the solution. 
There is a mass of $2.2\times 10^{-5}$ Jupiter masses
in the Roche lobe and the accretion rate, i.e.\ the mass taken off the grid,
is $1.5\times 10^{-4}$ Jupiter masses per orbit. 
The time that mass resides in the Roche lobe in this model is
only about a tenth of an orbital period.
The non-zero accretion rate is interesting because we include no
viscosity in our model. One might expect that planets opening a gap
in an inviscid disk could not accrete any more mass. 
There is the possibility that the planet-disk torques
together with the induced Reynolds stresses provide enough angular momentum 
transport to feed the planet via the circumplanetary accretion disk. 
Nevertheless, to clarify this issue of the accretion rate onto the planet the
simulations will have to be continued for many
more orbits at high resolution.
{ It may still be possible that accretion will stop after a
thousand orbits, which is the case in non-viscous
2D simulations of gap opening.}

We also calculate the gravitational torques acting on the planet
and translate them into the typical radial migration time  \citep{2001ApJ...547..457K}.
We measure the torques in three ways: 
excluding the mass contained in the inner half { Hill radius of the Roche lobe}
    (in the following the dashed curves), excluding all the { Roche lobe} 
(solid curves) and
excluding 1.5 times the Hill radius (dotted curves).
\citet{2003MNRAS.341..213B} have used this method to
give error bars to the measurements of the torques from the numerical simulations. This procedure is also very useful to study
the relative importance of the torques generated from gas close 
to the planet. 
In Fig.~\ref{fig:modCDI_torques.ps}
we see that the torques vary over time but the mean drift rate
corresponds to the $10^5$ yrs observed in other simulations and which fit
the results from linear theory \citep{1980ApJ...241..425G,
1997Icar..126..261W,2002ApJ...565.1257T}.
It is remarkable how close the three measurements of the torques are at
each time. This is a consequence of the fact that we take the mass out
of a significant part of the Roche lobe producing a very clean gap.
There is not enough matter left, to exert a torque on the 
planet. This changes in the models that do not treat mass accretion onto the
planet ($q_\mathrm{acc}=0.0$) or remove gas from a smaller part of the {Roche lobe} ($q_\mathrm{acc}=0.2$).
\subsubsection{Model CD4I}
\begin{figure}
\resizebox{\hsize}{!}
{\includegraphics{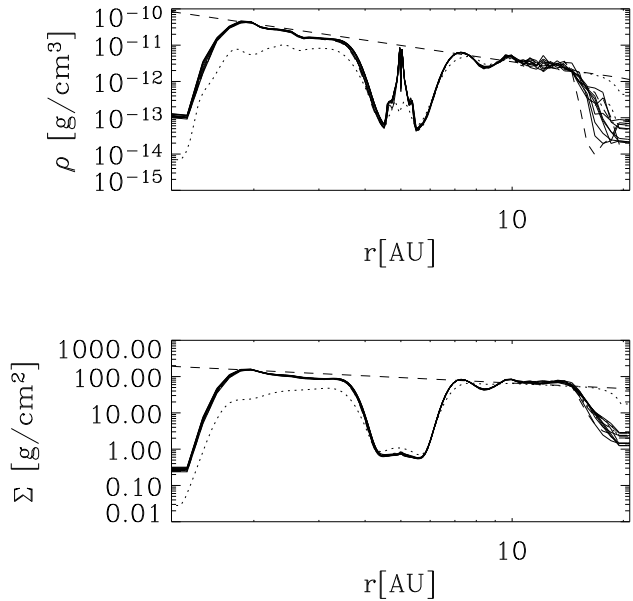}}
\caption{\label{fig:modCD4I_pstats.ps} 
Model CD4I:  Snapshots taken each orbit from t=71 to t=81 orbits. 
The radial slices of the density are taken in the midplane at the location of the planet. 
The surface density is azimuthally averaged.
The dotted line indicates the values for model CDI for comparison.
}
\end{figure}
We also have performed simulations with the same parameters as for model CDI
but with a higher resolution due to logarithmic concentration of grid cells
around the planet. Now the gravitational smoothing length
 is set to a value of only 20\% of the Hill radius
 ($q_\mathrm{g} = 0.2$), and the accretion radius to the same small
 region
 ($q_\mathrm{acc} = 0.2$).
The radial slices (all orbits from t=71 to t=81 orbits) through the planet
(see Fig.~\ref{fig:modCD4I_pstats.ps}) show that more mass
has accumulated in the Roche lobe than in model CDI (the values from model CDI are given by the dotted line).
This result can be expected because of the deeper gravitational potential of the planet;
the increase of density towards the planet is clearly visible.
The inner disk is less depleted than in model CDI, but
this indicates the earlier evolutionary step (81 orbits for model CD4I vs. 184 orbits for model CDI).
This model apparently loses more mass at the outer boundary, but this
is mainly due to the very coarse grid in combination with the open
boundary conditions there.

Nevertheless, the overall structure of the gap has not changed substantially. 
This model takes about 5 times more computational time than
the low resolution case. This is mainly due to the fine resolution
in the azimuthal direction, i.e.\ the Courant condition for the 
azimuthal velocity component.
Table 1 shows that there is a little more mass
accumulating in the Roche lobe (which was empty initially) than
in the low resolution case.
The torques are now strongly fluctuating while the average values are 
very similar to those of model CDI (see 
Table 1). The fluctuations are so strong that they even change their sign.
Again the time averaged torques are { insensitive} to the radius around the
planet, which has been excluded from the calculation of the torque.
We suspect that the oscillations might be an effect of the high density
concentration near the planet (see Fig.~\ref{fig:modCD4I_pstats.ps}), because
small density asymmetries induce large torque variation.
Similar findings have been obtained for nested grid simulations by  
\citet{2002A&A...385..647D}.

In future work we will return to mesh refinement around the planet to study these effects in more detail. 
\subsection{Radiative models}
\subsubsection{Model CD}
%
\begin{figure}
\resizebox{\hsize}{!}
{\includegraphics{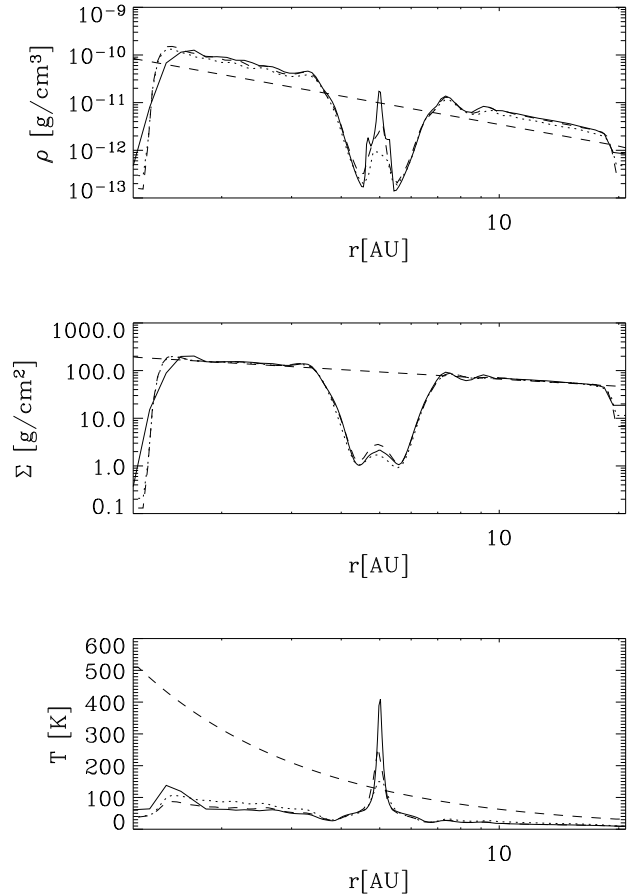}}
\caption{\label{fig:modCD4_pstats.ps} 
Model CD4: Snapshot at t=5 orbits
after the refinement of the grid. 
Radial slices of density and temperature are taken in the midplane at the location of the planet,
while surface density is the azimuthal average.
(solid line = Model CD4; dotted line = Model CD; dashed line = Model CDS).
}
\end{figure}
\begin{figure}
\resizebox{\hsize}{!}
{\includegraphics{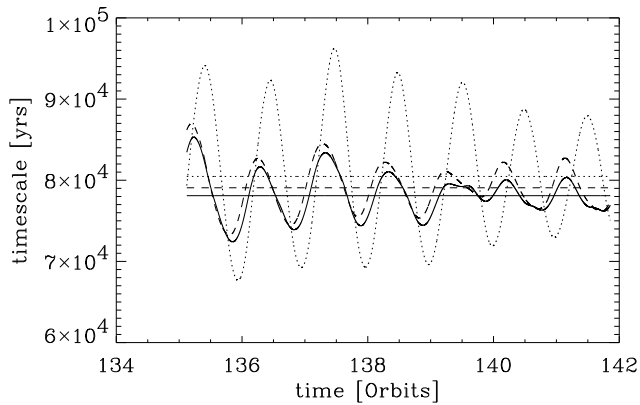}}
\caption{\label{fig:modCD_torques.ps} 
Model CD: Evolution of the planet migration rate in terms
of migration time. The solid curve is based on the 
torques excluding the Roche lobe and the solid horizontal
line is the time average. In addition we plot the torques based
on an exclusion of 0.5 Hill radii (dotted lines) and 1.5 Hill radii (dashed lines).
The horizontal lines refer to the time averaged values.
}
\end{figure}
%
The isothermal approach used in the previous models is
a strong simplification of the thermodynamics of the disk near the
planet and may not be well suited for studying the gap opening problem. 
For the radiative model CD
we keep all parameters identical to model CDI, but use the equation for
internal energy including flux--limited radiation transport, see Eq.~\ref{eq:energy}.
Again, mass is removed from within the Roche lobe, which
means that energy is taken out without releasing its
accretion energy (this will be treated differently in models DR and DR4).
The temperature and density structure of this model are
displayed together with other models in Fig.~\ref{fig:modCD4_pstats.ps} with dotted lines. 
In contrast to model CDI the disk has
cooled dramatically because there is no internal heating in our inviscid
model, while the region around the
planet has become slightly hotter than in the isothermal 
gap case (Fig.~\ref{fig:modCD4_pstats.ps}).
The rest of the gap is much cooler. 
There is also three times more mass piling up in the Roche lobe 
than in Model CDI (see also Table 1). This is a direct consequence of
the radiative cooling. If the matter within the Roche lobe cools down
it will contract leading to a higher density. 
The mass accretion rate onto the planet is roughly three to four times
larger than for the corresponding isothermal model.
Thus, there is more mass stored in the Roche lobe, which
also influences the torque (see Fig.~\ref{fig:modCD_torques.ps}).
While the overall average migration time is only 10\% less than
in model CDI, there are now variations between the torques measured
by excluding more or less mass from the {Roche lobe}.
Excluding either the entire {Roche lobe} or even 1.5 times the {Hill radius} gives
very similar results, while excluding only 0.5 of the {Hill radius} doubles the
amplitude of the fluctuations.

This is the first hint that radiative cooling changes the structure
of matter in the Roche lobe and consequently the resulting torques on the planet.
\subsubsection{Model CDS}
The following model is identical to the previous one but
this time mass is taken out of only 20\% of the Hill radius.
Thus we can study the effect of the arbitrary size of this
mass removal zone. As a result more mass accumulates in the 
Roche lobe close to the planet and the temperature
around the planet is even higher 
(Fig.~\ref{fig:modCD4_pstats.ps} dashed line).
Nevertheless, the model is already strongly limited by resolution
as 20 \% of the Hill radius means only two grid cells. 

The accretion rate is also close to the one in 
model CD.
The average torques
are still the same as in the 
isothermal case, model CDI.
It also shows that the migration times are the longest if one excludes 
only half the Hill radius. This indicates the importance of the proper treatment of mass
in the Roche lobe and its potential capability of altering the radial drift rate.
This model is also used for comparison to model CD4 and CD
in Fig.~\ref{fig:modCD4_pstats.ps}.
\subsubsection{Model CD4}
Here, we increase the resolution and decrease the gravitational
smoothing again to 20\% of the Hill  radius (cf. model CD4I). The radius out of which mass may be taken is decreased in the same fashion.
Otherwise this model (see Fig.~\ref{fig:modCD4_pstats.ps}) is identical to 
models CD (dotted line) and CDS (dashed line).

The most striking effect of the higher resolution is the
higher temperature at the planet's location.
Temperatures three times higher than in model CD
are reached.

This can be explained by two effects which are:
1. A deeper potential well leading to stronger 
compression and release of accretion energy.
2. The much higher densities near the planet leading to higher optical
depth and lower cooling rates.

Again the accretion rate 
is similar to model CD and
the average torques  
are also not very different (see Table 1).
\subsection{Non-accreting models}
   In all of the previous models mass has been taken out from the grid
in the vicinity of the planet. In the isothermal models
this might be a tolerable simplification as the average
torques are not different between the high and low resolution cases.

 However, in the models with radiation transport one introduces a large error
if one simply takes mass off the grid: Internal energy is removed from the
simulation along with the mass, but the accretion
energy released during final accretion phase onto the planet is neglected.
One can now either construct a model for this process and 
re-adjust the energy content of the gas after some mass has been
taken away (see Models DR and DR4) or simply refrain from accreting mass and 
study the effects of the mass piling up. The first
approach has one more free parameter than the second,
which is the relative accretion rate (i.e.\ the
fraction of the mass removed from the grid per orbit). 

We will first study more closely the effect of taking mass arbitrarily
from the grid. Therefore, we go back to the
isothermal model CDI but do not remove any mass this time.
\subsubsection{Model CDNI}
In comparison to model CDI the lack of accretion leads to a
slightly higher
concentration of mass around the planet, 
similar to the models with a deeper potential (see Table 1). 
The mass in the Roche
lobe is now three times greater than in model CDI.

   The major effect of not taking mass off the grid and letting it pile 
up in the Roche lobe is the introduction of strong fluctuations in
the migration rate. 
The torques calculated including half of the Hill radius
especially showed large fluctuations.
Nevertheless, the time averaged migration rate is still dominated by
the structures outside the Roche lobe and thus are of the same magnitude as
in all the previous models (see Table 1).
\begin{figure*}
\resizebox{\hsize}{!}
{\includegraphics{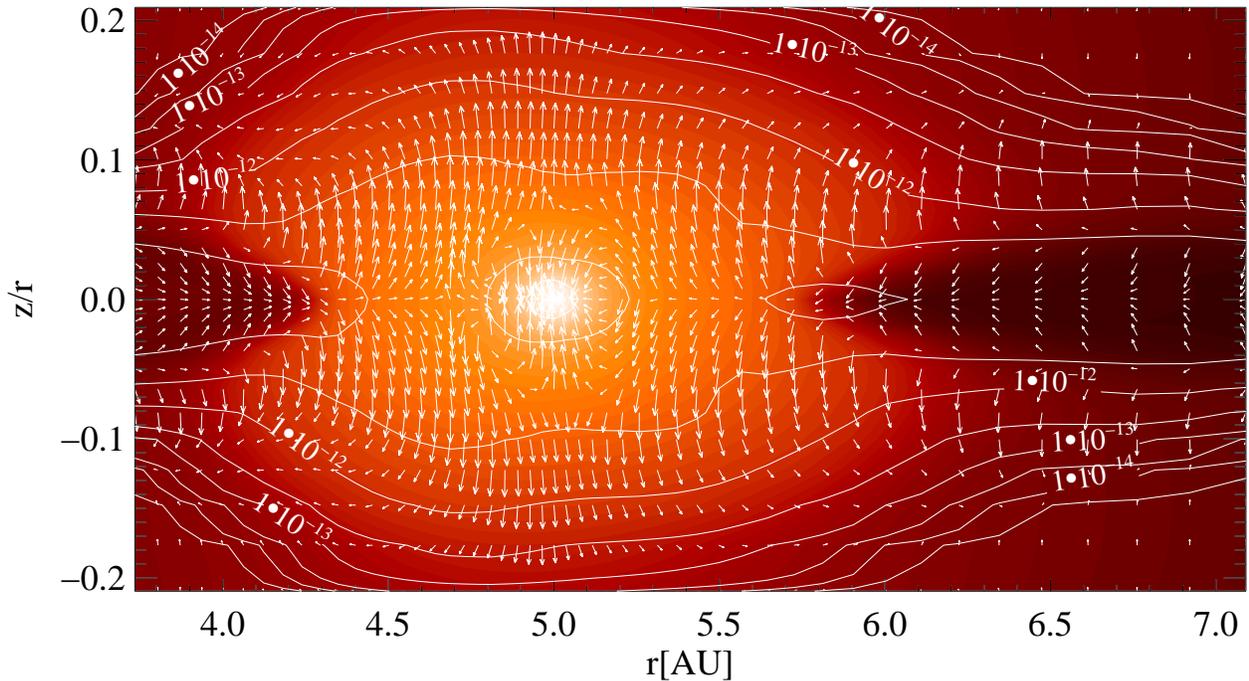}}
\caption{\label{fig:modDR4_ttxy2_flow.ps} 
Model DR4: Early stage of gap opening. Temperature in the r-$\theta$ plane of the protoplanetary disk
at the azimuthal location of the planet. This snapshot is taken
after 12 orbits, thus the gap is not completely empty yet. 
In this early phase the heating by the accretion onto the planet 
drives a strong convectional flow in the still {optically} thick disk
material above the planet.
{Brightness} is logarithmic temperature (lighter = warmer) ranging from $38$Kto $1060$K and vectors give the
logarithmic mass flux. We also plot iso-density lines with 
labels giving values in  g cm$^{-3}$. Density ranges from
$10^{15}$ g cm$^{-3}$ up to  $1.3 \times 10^{10}$ g cm$^{-3}$.
See the online edition for a color version of this plot.
}
\end{figure*}
\subsubsection{Model CDN and CDN4}
\begin{figure}
\resizebox{\hsize}{!}
{\includegraphics{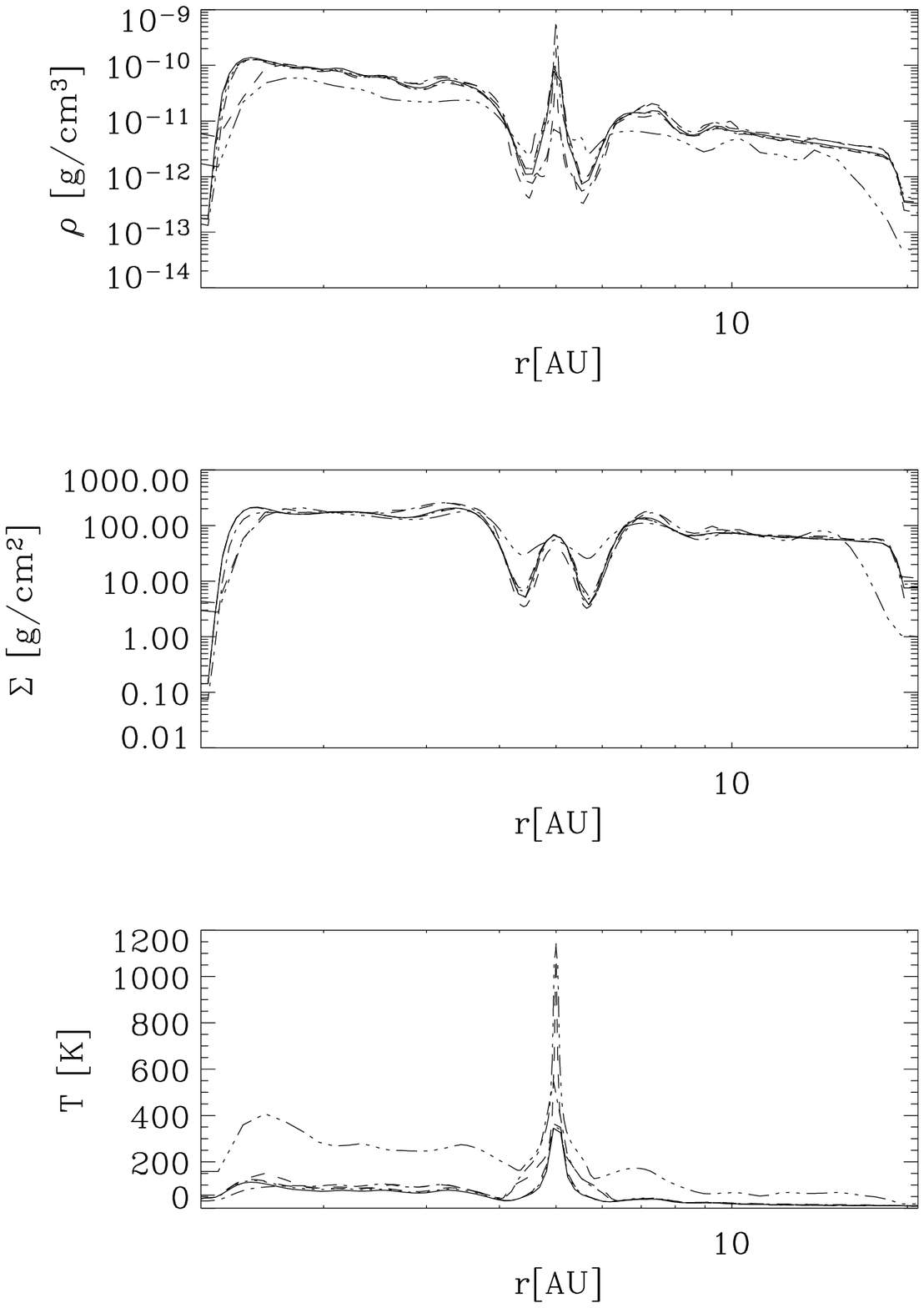}}
\caption{\label{fig:modDR_pstats.ps} 
{Model DR: Snapshot taken after t=121 orbits (solid line). 
Model DR4: after t=141 orbits (dotted line). 
Model DN: after t=121 orbits (dashed line). 
Model DN4: Snapshot taken after t=55 orbits (dash dotted line). 
Radial slices of density and temperature are taken in the midplane at the location of the planet, while surface density is the azimuthal average.}
}
\end{figure}
In these models we start with a gap, do not
remove any mass and perform the full radiation 
step. 
{Models starting with a gap eventually converge to models
without an initial gap.
This convergence can be demonstrated for the low resolution
models (see Table~1), while
the high resolution models are too expensive in computational
effort to show this in a reasonable computation time.}
The general trend can be seen in Table~1. Density and
temperature increase with the resolution of the Roche lobe while the 
torques are not much {different in the high and low resolution cases}.
\subsubsection{Model DN and DN4}
Instead of beginning with an artificial gap and then slowly filling 
the Roche lobe to accumulate mass around the
planet, we start here with an unperturbed axisymmetric
disk without a gap. 
The main result is that now a pressure supported atmosphere 
builds up around the planet, which becomes very hot. Both models
become stationary and we
observe maximum temperatures of about 600 K for model DN and 1150 K for the
high resolution case DN4 (see Table 1).
The radial profiles for these models along with the models DR and DR4
are displayed in Fig.~\ref{fig:modDR_pstats.ps}.
We remove no mass from
the grid but treat the full radiative dynamics.
The fact that the temperature is constant can be interpreted
with a self regulated accretion process into the Roche lobe.
The mass accretion rate balances the contraction of the cooling
planetary envelope and this contraction releases potential 
energy in the planetary gravitational field.
Thus we have a balance between mass flux 
and radiation flux over the surface of the Roche lobe.
The equilibrium
between cooling and accretion now yields a temperature that only 
depends on the depth of the potential well. This leads us to assume
that the temperatures will be even higher if we can avoid the 
artificial smoothing of the potential well. This is partly tested
by model DN4 with higher resolution and a smaller smoothing parameter (see Table 1).
Interestingly there is also a good agreement in temperature structure with the models that 
conserve total energy during accretion (see Fig.~\ref{fig:modDR_pstats.ps}). 
%
\begin{figure}
\resizebox{\hsize}{!}
{\includegraphics{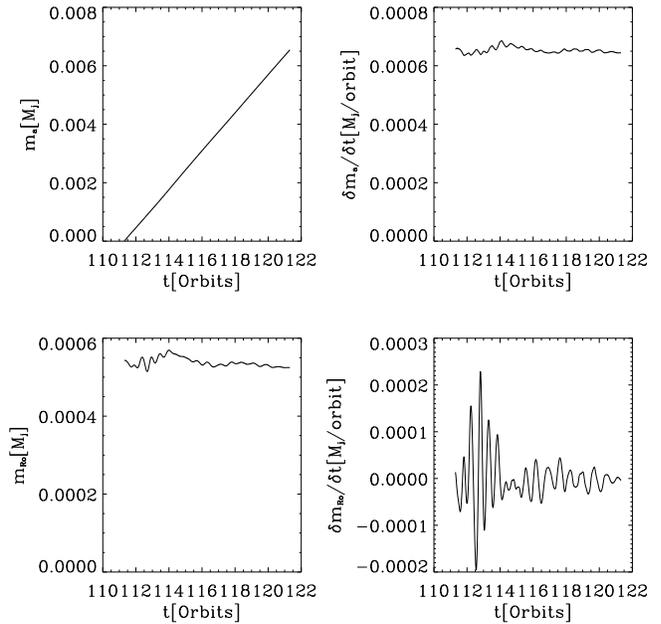}}
\caption{\label{fig:modDR_planetmass2.ps} 
Model DR: (upper row) accreted mass $m_a$ and accretion rate $\dot m_a$
onto the planet over time; (lower row) evolution of the mass contained in the Roche lobe $m_{\rm Ro}$ and
mass change in the Roche lobe $\dot m_{\rm Ro}$ in units of Jupiter masses and orbital periods of the planet.
}
\end{figure}
\begin{figure}
\resizebox{\hsize}{!}
{\includegraphics{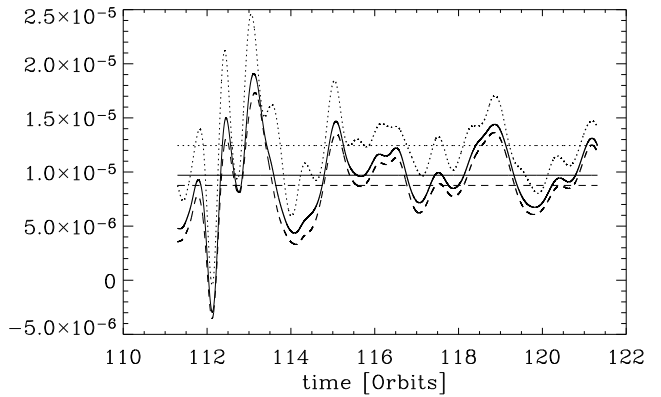}}
\caption{\label{fig:modDR_torques.ps} 
Model DR: Evolution of planet migration rate in terms
of migration time. The solid curve is based on the 
torques excluding the Roche lobe and the solid horizontal
line is the time average. In addition we plot the torques based
on an exclusion of 0.5 {Hill radii} (dotted lines) and 1.5 {Hill radii} (dashed lines), plus the time averaged values.
}
\end{figure}
\subsection{Accreting and energy conserving models}
\subsubsection{Model DR and DR4}
This model is similar to model CD in the sense that
we accrete mass from the Roche lobe. But this time it is
started without a gap and we conserve the total energy
locally in the grid cells where the gas is removed from the grid. 
Hence, the kinetic energy
and the potential energy that can be released
when the gas is accreted onto the planet are added locally 
to the internal energy.
The radius of such a young planet probably will be about two times the current size of
Jupiter (Bodenheimer, priv. communication).

First we show the flow situation in model DR4 after 12 orbits past the
initial set up. The gap has not cleared completely 
and the heating by accretion onto the planet drives 
strong vertical convection in the optically thick disk material 
above the planet (see Fig.~\ref{fig:modDR4_ttxy2_flow.ps}).
This "fountain" flow is subsonic and will not escape from the disk 
because the velocity is too low in comparison to the escape velocity.
Nevertheless, the effect might be very important for observations because
even embedded planets still too small to open a gap 
will probably produce this effect, which shows as an extended 
bubble of hot gas to the planet thus raising the local
pressure scale height in the disk.
 
Model DR approaches a quasi steady state after 100 orbits
(see Fig.~\ref{fig:modDR_pstats.ps}), which can also
be inferred from the almost constant Roche lobe mass of $5\times 10^{-4} M_{Jup}$ and 
the accretion rate onto the planet of $6\times 10^{-4} M_{Jup}$ per orbit
(see Fig.~\ref{fig:modDR_planetmass2.ps}). Nevertheless, there are also short time scale
fluctuations in the Roche lobe mass, which again show up in the fluctuations
of the torques (see Fig.~\ref{fig:modDR_torques.ps}).
Based on non-viscous 2D runs we know that it will take
up to a thousand orbits before the gap is so deep that no accretion onto
the planet occurs any longer. In 3D with radiation transport we presently
cannot afford such long runs.
The resulting maximal temperature in the Roche lobe is 800~K for model DR and 1500~K for model DR4. 
A finer grid resolution might 
even lead to higher temperatures.

\begin{figure}
\resizebox{\hsize}{!}
{\includegraphics{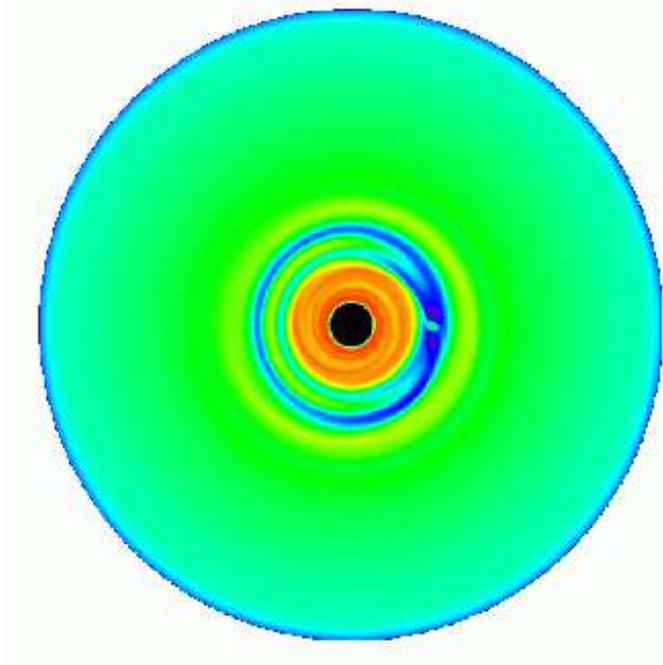}}
\caption{\label{fig:modDR_density_color.ps} 
Model DR: Density in the midplane of the protoplanetary disk after 121 orbits.
Dark means low, light means high density. The density ranges 
from $2.4 \times 10^{-13}{\rm g}\,{\rm cm}^{-3}$ to $8.3 \times 10^{-11}{\rm g}\,{\rm cm}^{-3}$.
See the online edition for a color version of this plot.
}
\end{figure}
%
\begin{figure}
\resizebox{\hsize}{!}
{\includegraphics{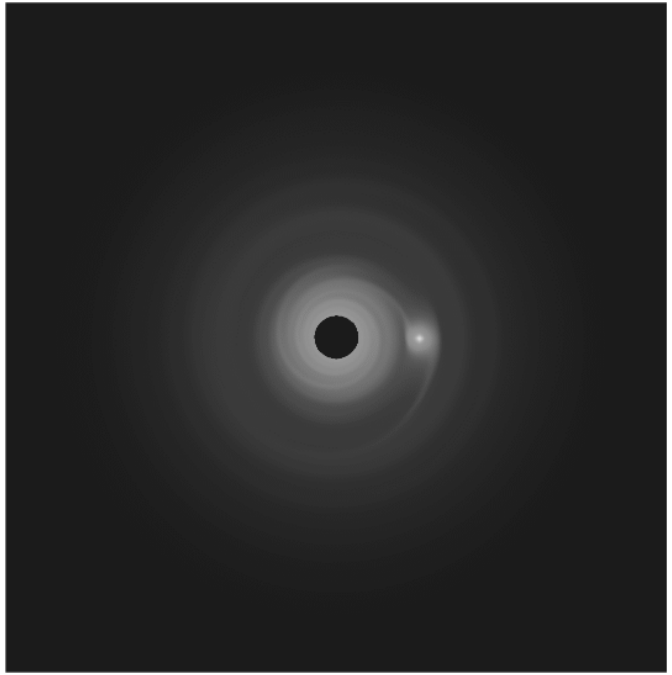}}
\caption{\label{fig:modDR_temperature_color.ps} 
Model DR: Temperature in the midplane of the protoplanetary disk after 121 orbits. Brightness corresponds to the logarithm of temperature.
Black indicates $10$K and white $800$K. 
See the online edition for a color version of this plot.
}
\end{figure}
\begin{figure}
\resizebox{\hsize}{!}
{\includegraphics{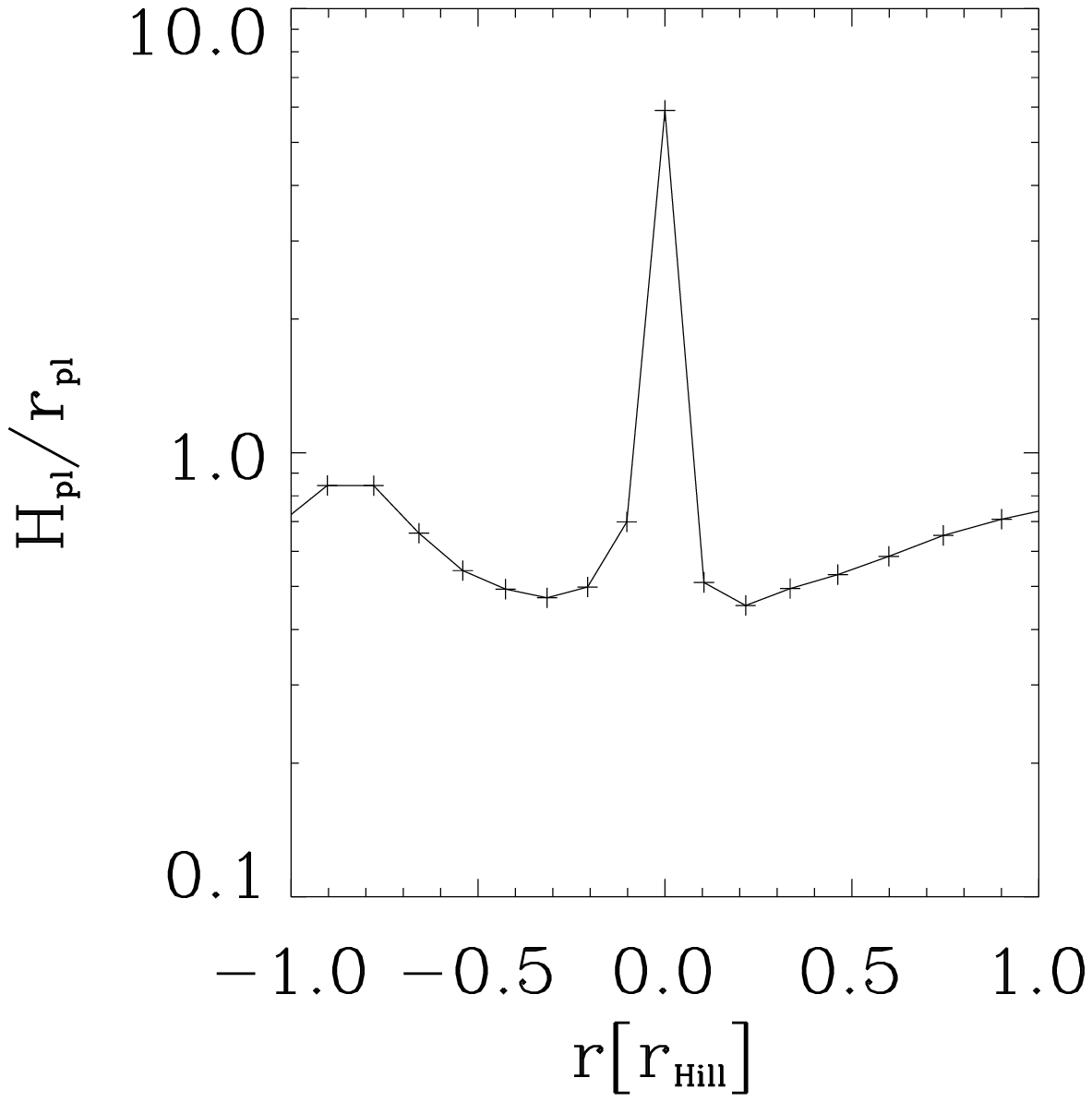}}
\caption{\label{fig:rstathor.ps} 
Model DR4: Pressure scale height in the midplane of the 
circumplanetary cloud after 141 orbits. Radial location is given in 
units of Hill radii.
}
\end{figure}

For model DR we also display the density structure that results in the midplane
 (see Fig.~\ref{fig:modDR_density_color.ps})  after 122 orbits. The gap has 
clearly formed but still mass is left in the {horseshoe} orbit. This mass will
decrease gradually over time. The density structure does not change significantly from
any isothermal simulation, but the temperature of course does 
(see Fig.~\ref{fig:modDR_temperature_color.ps}). The planet is the
hottest spot in the simulation. 
{We adopted a grey approximation for the radiation transport, thus continuum (wavelength dependent) radiation transport simulations (using our temperature and density structure)
will show whether this hot circumplanetary gas--dust mixture will be observable.} 
The extent of the warm gas around the planet
as well as the two mass feeding streams of warm gas hitting the circumplanetary
cloud are clearly visible now. We do not resolve any circumplanetary accretion disk yet, for two
reasons: First our resolution is too low and second the pressure dominates
the hot gas around the planet strongly enough so that no Keplerian disk forms, and one is
left with a pressure--supported planetary envelope filling
most of the Roche lobe.

In Fig.~\ref{fig:rstathor.ps} we display the pressure scale height of the gas 
in the midplane around the planet. In the circumplanetary ``disk'' 
$H_p/R_p$ is the inverse Mach number $1/M = c_s / v_k$ of the flow around the planet.
Due to the high temperatures the relative vertical thickness is about 0.5,
which is larger than those values
typically used in thin Keplerian accretion disks with $H/R < 0.1$. 
In the case of $H_p/R_p = 0.5$ this structure is better described
by a rotating envelope which additionally shows
strong deviations from a Keplerian rotation law of about $50 \%$.

\begin{figure*}
\resizebox{\hsize}{!}
{\includegraphics{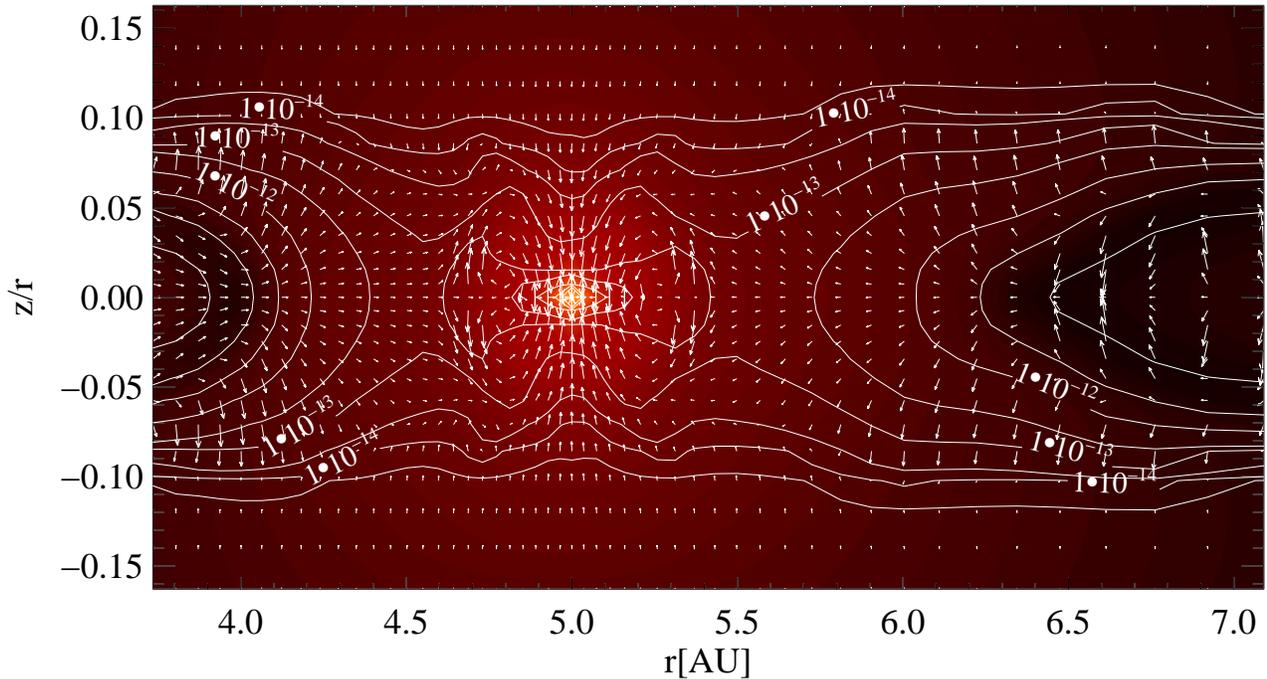}}
\caption{\label{fig:modDR4_ttxy2_global.ps} 
Model DR4: Temperature in the r-$\theta$ plane of the protoplanetary disk
at the azimuthal location of the planet after 141 orbits.
Brightness is logarithmic temperature (lighter = warmer) between $30$K and $1500$K, contours
are equi-density lines (in g cm$^{-3}$) and vectors give the logarithmic mass flux.
See the online edition for a color version of this plot.
}
\end{figure*}

For model DR4 we display some details of the flow onto the planet
(see Fig.~\ref{fig:modDR4_ttxy2_global.ps}).
We plot the temperature in the $r-\theta$ plane of the protoplanetary disk
at the azimuthal location of the planet. The temperature is hottest around the
planet and the gas in the gap is also heated by the accretion luminosity of
the planet.
The flow indicates that accretion mainly occurs via the poles of the 
planet and that there is convection in the envelope around the planet.
{We can see no inflow along the equatorial plane. The mass
falling onto the central region around the planet originates only in part
from the gap material above the planet. The largest contribution in 
mass is via the convective flow in the upper part of the circumplanetary cloud.}

This simulation also suggests that at least at this stage of gap opening
and planet disk interaction there is no formation of moons from
the circumplanetary material possible. The reason is simple: with $H_p/R_p = 0.5$
and the resulting strongly sub-Keplerian rotation, 
all solid material will quickly 
rain down onto the planet. The maximum radial drift velocity is 
given by the difference between the actual azimuthal velocity and
and the Keplerian value \citep{1977MNRAS.180...57W}. Thus before 
solid material has orbited the planet once, it will fall onto
 the planet. The fastest drifting solids are given by unity in the 
parameter $\Omega_p \tau_{\rm f} =1$, with $\Omega_p$ the Keplerian 
frequency and $\tau_{\rm f}$ the friction time of the solid body.
For the physical conditions in our model DR4 (e.g.\ density and temperature)
the fastest drifting objects would be 1 meter sized boulders,
which drift into Jupiter in about a month if released at 0.1 Roche radii.
Thus the drift problem for the formation of moons from circumplanetary 
material is even more crucial than in the case of planet formation.
We conclude that at least for the time when Jupiter is accreting its mass
there can be no formation of a satellite system.
\section{Conclusions}
We have performed full 3D radiation hydrodynamical simulations of embedded protoplanets
in disks, and compare the results in detail to the standard isothermal approach.

Mean torques and migration rates are not strongly affected by the treatment of the
thermodynamics in the case of Jupiter mass planets. This might change for lower
mass planets, which are more embedded in the disk. The fluctuations of the torques
on the other hand are much stronger, in particular in higher resolution cases.
Similar effects have been observed in
high resolution nested grid simulations \citep{2002A&A...385..647D, 2003ApJ...586..540D} and also
MHD simulations of planet-disk interactions \citep{2003MNRAS.339..983P, 2003MNRAS.339..993N,
2003ApJ...589..543W}. In some cases the torques even change their sign for a short period.

We find that planets are most likely to form a circumplanetary pressure--supported envelope
rather than an accretion disk around them, with strong convective vertical flows.
The relative pressure scale height in the circumplanetary material is at least $0.5$, 
in which case the approximations for a thin
Keplerian accretion disk are no longer valid. What results is a cloud which rotates at
only 50 percent of the Keplerian value. 

This slow rotation of the circumplanetary gas makes it impossible 
to form the moons of Jupiter from circumplanetary material as the
solid content will rain down
the planet in less than an orbit around Jupiter.

How does this thick and slowly rotating
envelope behave with respect to magneto-centrifugal
wind as discussed for a circumplanetary disk by \citet{2003A&A...411..623F}?
The high thermal pressure could be beneficial for launching 
the magneto-centrifugal wind, but on the other hand we observe a strong 
mass infall onto 
the poles of the planet and the surface of the circumplanetary cloud, 
thus a wind might be suppressed.
Detailed investigations on that issue are needed.

The deviation from symmetry with respect to the midplane is very small in
our simulations. Some convection cells show eventual overshooting
across the midplane but in general,
treating only one hemisphere would have led to the same results.

We observe strong vertical flows in the early gap opening phase.
Considering the entire mass accretion phase of the young planet
starting from a few earth masses \citep{1996Icar..124...62P}, one finds that these vertical 
fountains may last for hundred-thousands of years before the planet opens its gap, making the effect
clearly relevant for observations.
As a result there would be
locally a stronger flaring of the disk, and more radiation
could interact with the small dust grains in the surface layer 
above the planet.
Future instruments like SIM should be able to observe this asymmetry 
in the scattered light from disks in which planets are forming (Geoffrey Bryden personal communication). {On the other hand, the effect might be weaker if we allow
for a gradual buildup of the planet. This has to be investigated
in future simulations.}

We do not observe the formation of 
vortices at the edges of the gap, which can often be found in the equivalent
2D simulations \citep{1999MNRAS.303..696K, 2003ApJ...596L..91K}.
Further investigations will have to show whether this is related to the 3D nature
of our simulations or a consequence of the comparable low resolution 
of the grid.

Finally, we speculate that the extended warm gas and dust which fills most of the Roche lobe
should be detectable in the far infrared. Investigations such as by \citet{WolfDAngelo2005} indicate this
already for 2D flat simulations that include radiative cooling. 
Similar studies using 3D density and temperature distributions
will follow in the near future.
%
%
\begin{acknowledgements}
The authors wish to thank Geoff Bryden for fruitful discussions.
This work was sponsored by
the European Community's Research Training Networks Programme 
{\it The Origin of Planetary Systems} under contract HPRN-CT-2002-00308.
\end{acknowledgements}

\bibliographystyle{aa}

\begin{longtable}{l|l|l|l|l|r|l|l|l|r|l}
\hline
\hline
{name}  & {grid}                    &{EOS}& {$q_\mathrm{g}$;$q_\mathrm{a}$}&{ini}&{$t_\mathrm{max}$}&{$\dot M$}&{$M_\mathrm{Roche}$}&{$\rho_\mathrm{max}$}&{$T_\mathrm{max}$}&{$\tau_\mathrm{mig}$} \\
\hline
{CI}    &$100\times20\times 1  $    & iso  & 0.5   ; 0.5 &         & $300$ & -                  &     -             & $2.0\times10^{-14}$ &$130$& - \\  
{CDI}   &$100\times20\times 200$    & iso  & 0.5   ; 0.5 & G       & $184$ & $1.5\times10^{-4}$ & $2.2\times10^{-5}$& $3.0\times10^{-13}$ &$130$& $8.9\times10^{4}$  \\  
{CD4I}  &$100\times24\times 200$ lc & iso  & 0.2   ; 0.2 & G       & $81$  & $1.2\times10^{-4}$ & $3.2\times10^{-5}$& $0.9\times10^{-11}$ &$130$& $1.0\times10^{5}$          \\   
{CD}    &$100\times20\times 200$    & rad  & 0.5   ; 0.5 & G       & $142$ & $6.0\times10^{-4}$ & $6.5\times10^{-5}$& $1.0\times10^{-12}$ &$150$& $7.8\times10^{4}$          \\     
{CDS}   &$100\times20\times 200$    & rad  & 0.2   ; 0.2 & G       & $141$ & $3.0\times10^{-4}$ & $1.3\times10^{-4}$& $2.0\times10^{-12}$ &$200$& $8.2\times10^{4}$  \\     
{CD4}   &$100\times24\times 200$ lc & rad  & 0.2   ; 0.2 & G       & $75$  & $4.0\times10^{-4}$ & $8.0\times10^{-5}$& $2.0\times10^{-11}$ &$480$& $8.0\times10^{4}$  \\ 
{CDNI}  &$100\times20\times 200$    & iso  & 0.5   ; 0   & G       & $98$  &    -               & $3.3\times10^{-4}$& $5.0\times10^{-11}$ &$130$& $8.3\times10^{4}$  \\ 
{CDN}   &$100\times20\times 200$    & rad  & 0.5   ; 0   & G       & $44$  &    -               & $7.5\times10^{-3}$& $6.0\times10^{-11}$ &$310$& $8.9\times10^{4}$  \\ 
{CDN4}  &$100\times24\times 200$ lc & rad  & 0.2   ; 0   & G       & $122$ &    -               & $3.6\times10^{-4}$& $4.0\times10^{-10}$ &$520$& $1.2\times10^{5}$  \\  
{DN}    &$100\times20\times 200$    & rad  & 0.5   ; 0   &         & $121$ &    -               & $1.3\times10^{-3}$& $7.0\times10^{-11}$ &$600$& $6.6\times10^{4}$  \\ 
{DN4}   &$100\times24\times 200$ lc & rad  & 0.2   ; 0   &         & $55$  &    -               & $1.7\times10^{-2}$& $2.0\times10^{-9} $ &$1150$&    \\  
{DR}    &$100\times25\times 200$    & rad+ & 0.2   ; 0.2 &         & $121$ & $6.5\times10^{-4}$ & $5.5\times10^{-4}$& $2.0\times10^{-11}$ &$800$& $1.0\times10^{5} $  \\ 
{DR4}   &$100\times25\times 200$ lc & rad+ & 0.1   ; 0.1 &         & $141$  & $3.5\times10^{-4}$ & $4.5\times10^{-4}$& $4.5\times10^{-10}$ &$1500$& $1.2\times10^{5} $    \\ 
\caption{\label{tab1} Parameters chosen in the different
simulations and their results. These parameters are the dimensioning of the grid ($n_r, n_\vartheta,n_\varphi$),
the kind of spacing (lc = logarithmically centered around the planet), 
the thermodynamics (iso = locally isothermal, rad = ideal gas plus radiation transport, rad+ = ideal gas,
radiation transport and conservation of total energy for gas accreted onto planet), the gravitational smoothing radius ($q_\mathrm{g}$ = fraction of the Hill radius over that smoothing is applied), the accretion radius ($q_\mathrm{acc}$ = fraction of the Roche lobe (respectively Hill radius) that mass is taken out of), the type of the initial model (G: starting with an
existing gap), and the total number of orbits calculated ($t_\mathrm{int}$). The results we give are the accretion rate onto the Planet ($\dot M$) in units of Jupiter masses per orbit, the mass contained in the Roche lobe ($M_\mathrm{Roche}$), the maximum density and temperature in the Roche lobe ($\rho_\mathrm{max}, T_\mathrm{max}$) and the migration time-scale that we determined ($\tau_\mathrm{mig}$).}
\end{longtable}

\end{document}